\documentclass[12pt]{spieman}
\usepackage{amsmath,amsfonts,amssymb}
\usepackage{graphicx}
\usepackage{lscape}
\usepackage{rotating}
\usepackage{setspace}
\usepackage{tocloft}
\usepackage[normalem]{ulem}
\usepackage{multirow}

\title{Assessing Phase Reconstruction Accuracy for Different Nonlinear Curvature Wavefront Sensor Configurations}

\author[a,*]{Stanimir Letchev}
\author[a,b]{Jonathan Crass}
\author[a]{Justin R. Crepp}

\affil[a]{University of Notre Dame, Department of Physics and Astronomy, 225 Nieuwland Science Hall, Notre Dame, Indiana, 46556-5670, United States}
\affil[b]{The Ohio State University, Department of Astronomy, 4055 McPherson Laboratory, 140 West 18th Avenue, Columbus, Ohio, 43210-1173, United States}

\cftpagenumbersoff{figure}
\cftpagenumbersoff{table} 
\begin{document} 
\maketitle

\begin{abstract}

    The nonlinear curvature wavefront sensor (nlCWFS) offers improved sensitivity for adaptive optics (AO) systems compared to existing wavefront sensors, such as the Shack-Hartmann. The nominal nlCWFS design uses a series of imaging planes offset from the pupil along the optical propagation axis as inputs to a numerically-iterative reconstruction algorithm. Research into the nlCWFS has assumed that the device uses four measurement planes configured symmetrically around the optical system pupil. This assumption is not strictly required. In this paper, we perform the first systematic exploration of the location, number, and spatial sampling of measurement planes for the nlCWFS. Our numerical simulations show that the original, symmetric four-plane configuration produces the most consistently accurate results in the shortest time over a broad range of seeing conditions. We find that the inner measurement planes should be situated past the Talbot distance corresponding to a spatial period of $r_0$. The outer planes should be large enough to fully capture field intensity and be situated beyond a distance corresponding to a Fresnel-number-scaled equivalent of $Z\approx50$ km for a $D=0.5$ m pupil with $\lambda=532$ nm. The minimum spatial sampling required for diffraction-limited performance is 4-5 pixels per $r_0$ as defined in the pupil plane. We find that neither three-plane nor five-plane configurations offer significant improvements compared to the original design. These results can impact future implementations of the nlCWFS by informing sensor design.
    
\end{abstract}


{\noindent \footnotesize\textbf{*}Stanimir Letchev,  \linkable{sletchev@nd.edu} }

\begin{spacing}{2}

\section{Introduction}\label{sec:intro}

The design of a wavefront sensor (WFS) involves a fundamental trade-off between sensitivity, dynamic range, and speed \cite{haffert_16}. Challenging applications for adaptive optics (AO) systems, such as high-contrast imaging\cite{chilcote_22}, single-mode fiber injection for spectroscopy\cite{crass2020} and interferometry\cite{chara}, laser communications\cite{Majumdar2005FreespaceLC}, directed energy\cite{merrit_spencer}, remote sensing\cite{rs12101659}, and space domain awareness\cite{hart2016new}, have requirements that demand high sensitivity (ability to correct at low photon levels), a large dynamic range for aberration correction (phase errors of many waves), and fast operating speeds (many kHz) simultaneously. Various types of curvature sensing have been proposed as a method to address this challenge \cite{guyon_10,mateen_11}.

Curvature WFSs work on the premise that, as a beam of light propagates, the local shape of the wavefront phase will focus or defocus light in the direction of the curvature. The resulting pattern appears as a bright region on one side of the optical system pupil (or focal plane), and a darker region on the opposite side. By measuring the change in intensity between planes on either side, the local curvature of the wavefront can be obtained and the phase of the overall aberration profile can be reconstructed. However, as the distance between the measurement planes and pupil (or focal) plane increases, the wavefront response becomes nonlinear, resulting in poor wavefront reconstruction if the original linear assumptions are maintained.\cite{guyon_05,guyon_08}

Conventional curvature WFSs have generally relied on linear reconstructors to allow for stable, closed-loop operation at speeds sufficient to compensate for atmospheric turbulence. However, to achieve both a large dynamic range and high sensitivity simultaneously, a WFS must sacrifice its linear response to wavefront distortions \cite{haffert_16}. A nonlinear WFS often requires the use of iterative numerical reconstructors to retrieve the wavefront phase. The iterative nature of these reconstructors results in longer reconstruction times and increased latency. However, increasing computational power, made available through parallel computing, graphics processing units (GPUs), and embedded systems, such as field programmable gate arrays (FPGAs) and application-specific integrated circuits (ASICs), now allows for the development of WFS architectures and real-time reconstruction algorithms that are inherently nonlinear.

In light of these considerations, the concept of a nonlinear curvature WFS (nlCWFS), originally proposed in 2009\cite{guyon_10}, has become increasingly feasible as a solution. The nlCWFS design uses multiple measurement plane images that are positioned away from the pupil plane along the optical axis to reconstruct the wavefront shape of an aberrated beam of incoming light. Practical implementations of the sensor have generated these measurement planes by placing beamsplitters or dichroics in a collimated beam and recording the generated channels onto a detector or detectors .\cite{mateen_11,crass_14,Crepp_20} Numerical simulations and laboratory experiments have indicated that the nlCWFS (also referred to as a Fresnel WFS in the literature) can achieve an order of magnitude higher sensitivity than the industry-standard Shack-Hartmann WFS (SHWFS) and deliver improved performance compared to the pyramid wavefront sensor.\cite{guyon_10,mateen_11} The nlCWFS further offers a large dynamic range for wavefront reconstruction and the ability to measure amplitude errors induced by scintillation. Such features make the nlCWFS an excellent candidate for AO systems operating in the presence of ``deep turbulence." \cite{watnik_gardner_18,Crepp_20} However, because of its nonlinearity, the nlCWFS must use an iterative reconstruction method, typically based on the Gerchberg-Saxton (GS) method, where light is numerically propagated between all its measurement planes and the pupil to attempt to reconstruct the pupil phase and amplitude.

The nlCWFS allows for the ability to adjust the location of the measurement planes along the optical axis to provide optimal performance; varying the defocus distances changes the device's sensitivity to reconstructing certain spatial frequency aberrations. Since the nlCWFS attempts to measure the effects of near-field diffraction, the optimal location of the measurement planes should be related to the characteristic distance over which phase errors at the pupil are converted into amplitude errors at the measurement planes. Thus, for optimal reconstruction, the measurement plane detectors should be placed along the optical axis where the visibility of fringes due to a periodic disturbance is relatively high. The Talbot distance, $Z_T$, quantifies this characteristic length scale over which monochromatic interference fringes repeat as a coherent beam propagates,

\begin{equation}
  Z_T=\frac{2 a^2}{\lambda},
  \label{eq_talbot_dist}
\end{equation}
where $a$ represents the spatial period of a periodic disturbance and $\lambda$ is wavelength \cite{wen_13}. 

In practice, optimal measurement plane positions often depend on a power spectrum of aberrations experienced by the optical system. For a continuous spectrum of spatial frequencies and infinite signal, an ideal nlCWFS would use an infinite number of measurement planes, to measure and reconstruct the entire three-dimensional spatial distribution of the field disturbance. A continuum of camera defocus distances is not practical to build, and would require impossibly-long reconstruction times due to the increased computational complexity of the reconstruction algorithm. Therefore, viable designs must determine an optimal number and location of measurement planes based on the aberrations experienced in a given application.

Previous numerical simulations and laboratory experiments of the nlCWFS have, by default, assumed four measurement planes, with two pairs located symmetrically on either side of the pupil plane: one pair closer to the pupil plane to sense higher spatial frequencies and another pair further away to sense lower spatial frequencies \cite{crass_14,mateen_15}. While this configuration might represent a natural extension of the original curvature WFS concept, it is not strictly required \cite{guyon_10}. A nlCWFS configuration with a different number of planes might be capable of accomplishing the same task. In addition, previous studies have typically determined optimal measurement plane distances through trial and error based on experimental results. These distances are then dependent on each individual system\cite{mateen_15,crass_14,Crepp_20}. Since the speed of algorithm convergence, accuracy of the wavefront reconstructor, and resulting noise characteristics of an AO system are affected by both the number and location of nlCWFS measurement planes, it is important to establish a better understanding of the optimal values of these parameters for a general system. 

In this paper, we explore the effects of varying the location and number of measurement planes (along the optical propagation axis), as well as the detector spatial sampling constraints (across the optical axis), in an attempt to optimize the nlCWFS configuration for applications involving Kolmogorov turbulence. In Section~\ref{sec:methods}, we describe the numerical physical optics methods used to perform the simulations for characterizing nlCWFS performance. In Section~\ref{sec:results}, we present simulations and analyses that quantify algorithm accuracy, speed, and convergence. The results are used to determine the optimal defocus distances of the four-plane configuration, to study the effects of changing the number and symmetry of the measurement planes, and to determine the effects of changing detector spatial sampling on algorithm convergence. Finally, in Section~\ref{sec:conclusions}, we present a summary of the results and discuss their implications.

\section{Numerical Methods}\label{sec:methods}

Due to the nonlinear optical propagation regime in which the nlCWFS operates and the large parameter space of plausible configurations (measurement distances, number of planes, spatial sampling, etc.), it is difficult to determine the optimal characteristics of the nlCWFS analytically. Instead, we have developed a series of simulations to model the sensor's performance numerically. This suite of scalar wave-optics simulations explores the nlCWFS's performance based on a variety of model parameters (pixel sampling, plane locations, convergence criteria) and observing conditions ($D/r_0$), where $D$ is the telescope diameter and $r_0$ is the Fried parameter. Custom MATLAB scripts were developed to model, sense, and reconstruct near-field (Fresnel) diffraction effects. These programs were further supplemented by commercially- and publicly-available MATLAB codes, as described below. The combined tools were used to generate simulated data and were also used as an integral part of the wavefront reconstruction algorithm. The simulations were performed with static aberrations without the presence of noise to test the accuracy and latency of reconstruction, rather than closed-loop performance. Potier et al. 2023 (submitted) present an error budget for the nlCWFS including noise sources. 

\subsection{Optical Propagation Method}\label{sec:prop_method}

The process for generating simulated images (\S \ref{sec:img_gen}) for the nlCWFS's reconstruction algorithm (\S \ref{sec:GS_alg}) relies on propagating the electric field between multiple measurement planes located outside of the optical system pupil. Therefore, it was necessary to choose a robust, yet fast, propagation algorithm for both simulations and phase recovery. Due to the distances, wavelength, and beam diameters used, all simulations were performed within the Fresnel (near-field) regime for coherent light propagation. As such, Fresnel propagation was used for all steps in the detector image generation process and reconstruction algorithm.

The propagation algorithms selected were based on those in Schmidt 2010\cite{schmidt_10}. Specifically, the angular spectrum method was used due to its ability to perform complete Fresnel propagation using a 1-to-1 pixel scale ratio, which best represents the physical setup of the detectors used for recording measurement plane images in the nlCWFS. In addition, this method is regarded as more accurate for near-field propagation compared to other techniques, such as two-step propagation.\cite{Papalexandris_00} Schmidt's angular spectrum code was slightly modified for efficiency, as many variables do not need re-calculation in each iterative loop. 

\subsection{Phase Aberrations and Image Generation}\label{sec:img_gen}

The simulated images that were used as inputs for the reconstruction algorithm are analogous to those captured by the detectors in the nlCWFS in a real-world scenario. The intensities at the measurement planes were generated by creating an electric field with a pupil phase aberration and uniform intensity. This electric field was propagated to the various measurement planes according to their distance from the pupil.

\begin{figure}
    \centering
    \includegraphics[width=\textwidth]{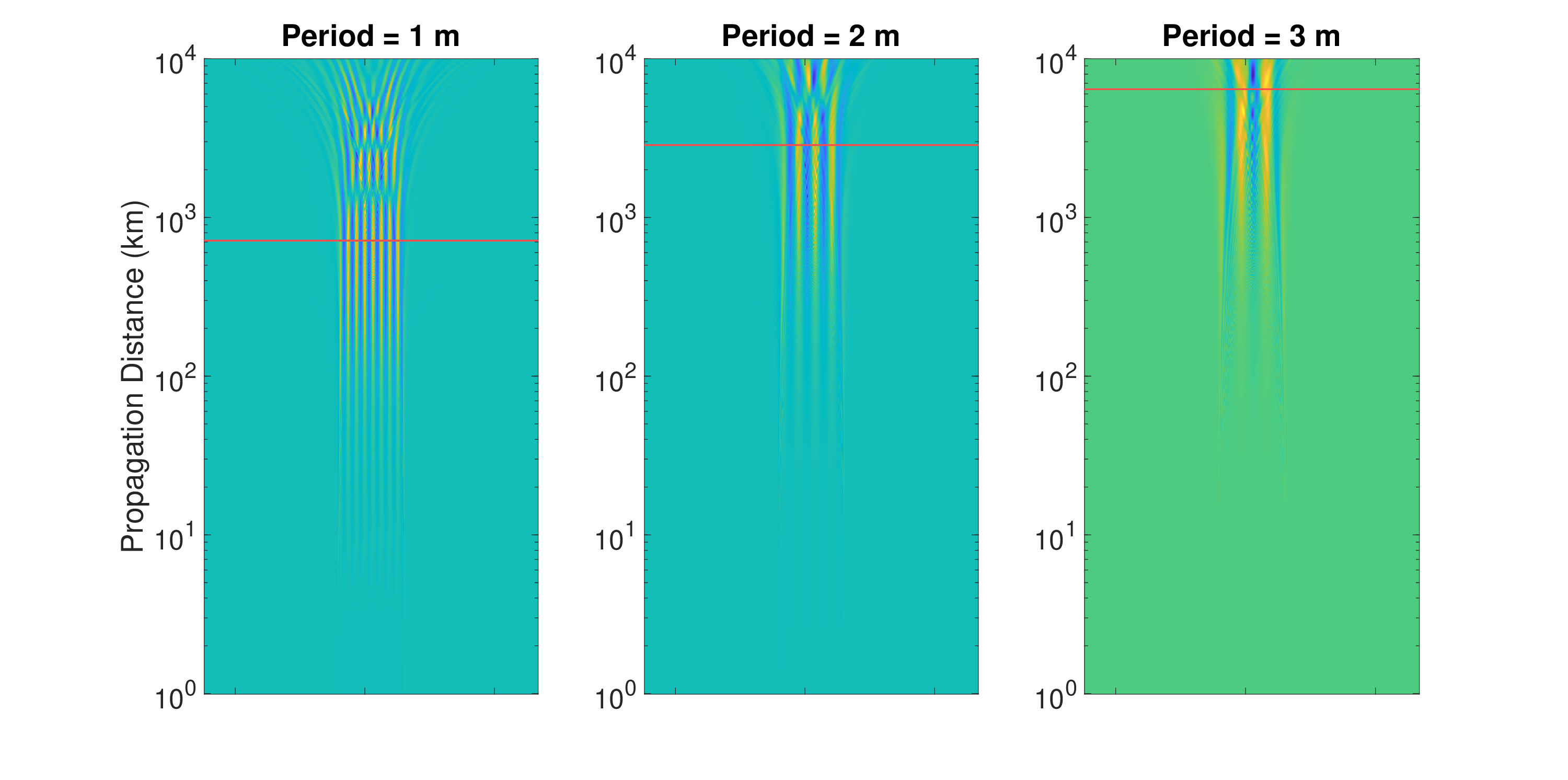}
    \caption{The differential intensity ($I_{+Z}-I_{-Z}$) from a propagation of a simple sinusoidal aberration with periods of 1, 2, and 3 m imposed on an 8 m pupil. The horizontal red lines correspond to the Talbot distance given by each spatial frequency. As expected, the strongest differential intensity occurs around the Talbot distance specified by the spatial frequency, verifying that our propagation algorithms are producing valid results.}
    \label{fig:sinusoidal_3_periods}
\end{figure}

Initially, a set of phase aberrations was created using a simple sinusoidal function to facilitate verification of the propagation code, replicating an analysis from Guyon 2008 which shows the differential intensity either side of the pupil (Figure \ref{fig:sinusoidal_3_periods}).\cite{guyon_08} All subsequent phase aberrations were created using the WaveProp \textit{kolmogphzscreen} function for atmospheric simulations.\cite{Brennan:16,Brennan:17} The phase masks corresponding to these aberrations were kept for comparison to ensure that residual phase errors were calculated based on the original phase values rather than on extracting a phase from the total electric field, as this would be subject to phase unwrapping errors.

The spatial sampling for atmospheric aberrations of $D/r_0{\le}8$ was set to correspond to a 64 pixel-diameter pupil, which ensured sufficient sampling to not limit reconstruction accuracy (see \S \ref{sec:sampling} for further details). Since stronger turbulence required additional sampling, for values of $D/r_0{>}8$, the pupil diameter was set to 256 pixels wide. To allow the phase generator to function correctly, ensure sufficient sampling in Fourier space for the reconstructor, and to minimize numerical ringing effects during propagation, the initial square pupil array was padded with zeros to a size of $2048{\times}2048$ for $D/r_0{\le}8$ and $4096{\times}4096$ for larger $D/r_0$ values. These arrays, once populated with an input phase in the central pupil region, were propagated to the measurement planes and recorded for use during reconstruction. For the $D/r_0{\le}8$ cases, recorded images were truncated to $1024{\times}1024$ to reduce runtime during reconstruction, while for larger $D/r_0$ values, the full array was used to avoid clipping effects (see \S \ref{sec:defocus_distance_analysis}).

The complex electric field, $U(x,y)$, at the pupil was created from the generated phase masks by combining each phase mask with a uniform electric field amplitude as follows:
\begin{equation}
\label{eq:electric_field_calculation}
    U(x,y)=\sqrt{\frac{n_{\rm photons}}{n_{\rm pixels} \; n_{\rm planes}}}e^{i\frac{2\pi}{\lambda}\phi(x,y)}
\end{equation}
where $\phi(x,y)$ is the generated phase, $n_{\rm photons}$ is the number of photons entering the optical system, $n_{\rm pixels}$ is the number of pixels illuminated at the pupil, and $n_{\rm planes}$ is the number of measurement planes into which light is divided.

\begin{figure}
    \centering
    \includegraphics[trim=0 0 0 0,clip,width=\textwidth]{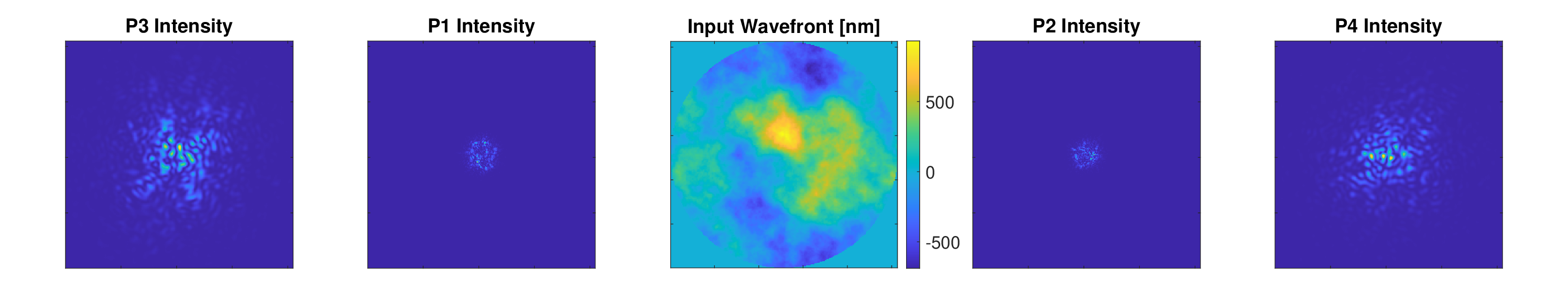}
    \includegraphics[trim=0 20 0 50,clip,width=\textwidth]{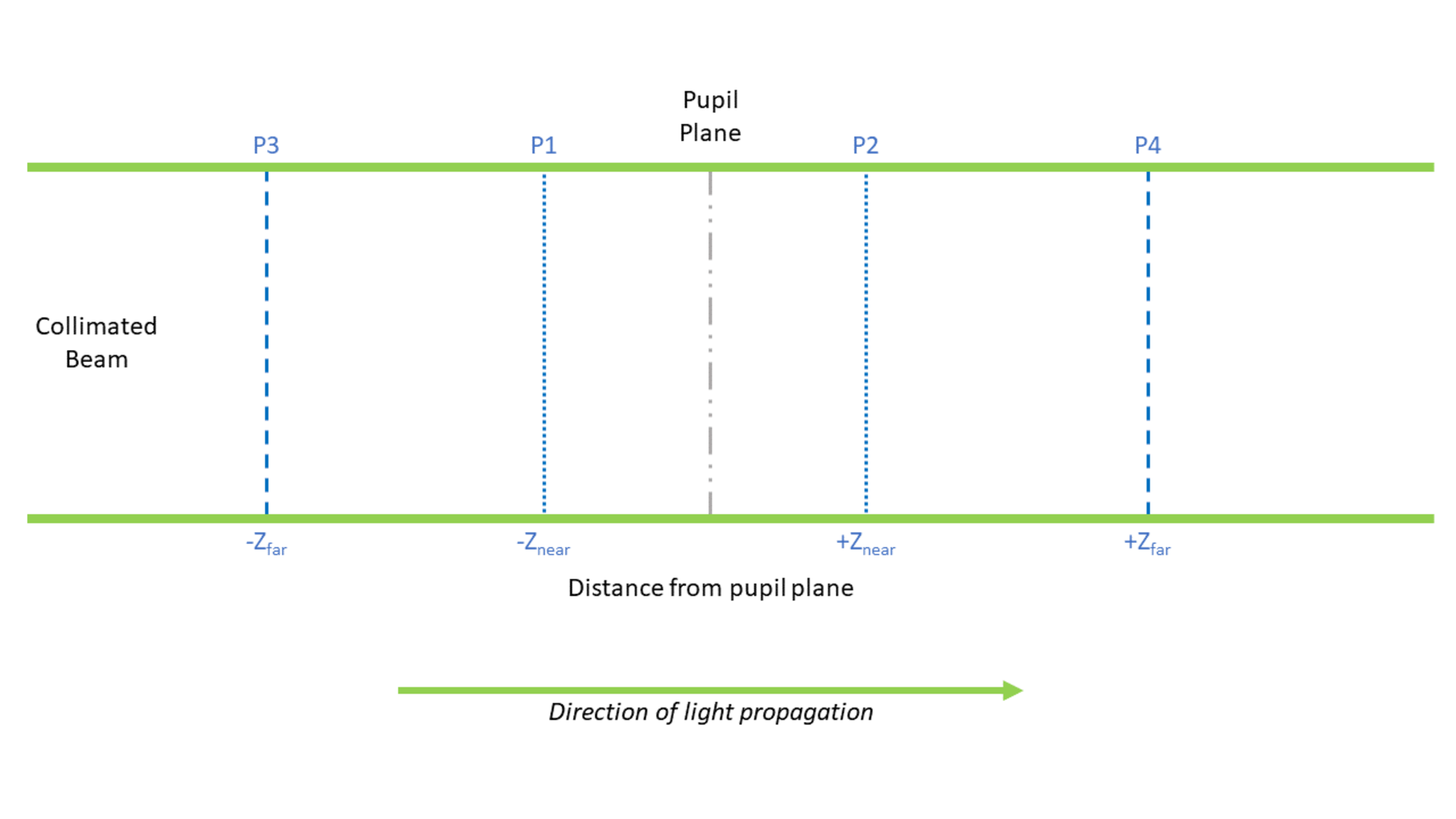}
    \caption{Visual representation of the plane configuration names and definitions used in this study. Figure adapted from Crass et al. 2014 \cite{crass_14}. The upper panels shows representative intensity distributions at the measurements planes for $D/r_0$ of 16 with a 0.5 m pupil ($Z_{near}$ = 9 km; $Z_{far}$ = 75 km).}
    \label{fig:plane_config}
\end{figure}

The generated electric field at the pupil was propagated to each measurement plane using the method outlined in $\S$\ref{sec:prop_method}. The planes were defined using the same notation as in Crass et al. 2014\cite{crass_14} (Figure \ref{fig:plane_config}). For the symmetric four-plane configuration, the main parameters that define the sensor are $Z_{\rm near}$ and $Z_{\rm far}$, which represent the distances from the pupil for the near and far planes, respectively. The simulated intensities at each plane, $I_n$, analogous to detector measurements in a real-world nlCWFS, were generated by:
\begin{equation}
\label{eq:defocus_plane_intensity}
    I_n(x,y)=U_n^*(x,y) \times U_n(x,y)
\end{equation}
where $U_n$ is the propagated electric field at the n\textsuperscript{th} measurement plane. These intensities were then used as input for the wavefront reconstruction process, examples of which are shown in Figure \ref{fig:plane_config}.

\begin{figure}
    \centering
    \includegraphics[width=\textwidth]{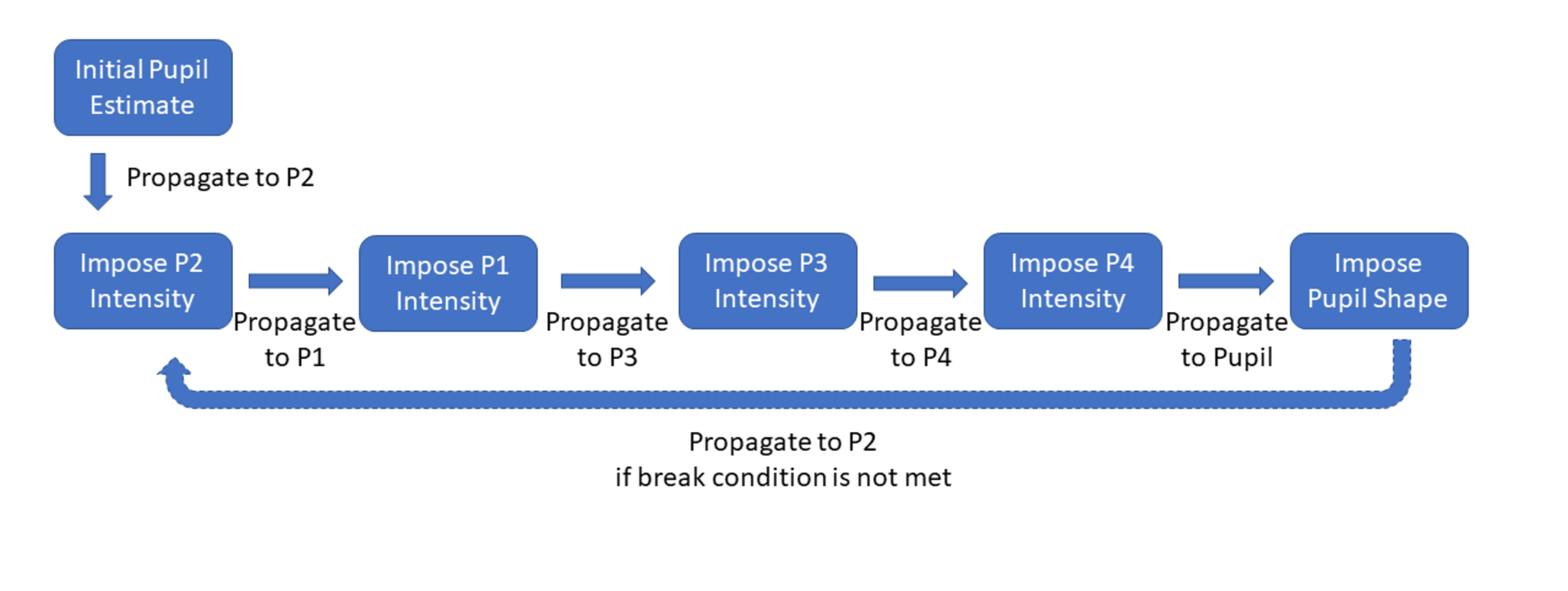}
    \caption{Visual representation of the modified GS algorithm used for wavefront reconstruction, showing the propagation order and steps. The dashed-border arrow represents where the loop would break when the convergence criterion is met. Figure adapted from Guyon 2010.\cite{guyon_10}}
    \label{fig:GS_Algorithm}
\end{figure}

\subsection{Modified Gerchberg-Saxton Algorithm}\label{sec:GS_alg}

The reconstruction method used for simulating the nlCWFS is based on the modified GS algorithm outlined in Guyon 2010\cite{guyon_10}. The electric field is propagated from one measurement plane to another. The intensity measurement at each location is imposed onto the field by replacing the amplitude with the square-root of the intensity. The electric field is eventually propagated back to the pupil as shown in Figure \ref{fig:GS_Algorithm}. Field points located outside of the pupil diameter are set to zero to minimize numerical noise, since no light should arise from outside of that boundary. This modified algorithm differs from the original GS method in that there are multiple plane propagations in a loop iteration, instead of a pair of propagations back and forth between the source and image planes.

As the algorithm is iterative, a criterion for when it has reached a solution is required. In our simulations, the convergence condition was defined as a lower limit to the RMS difference in wavefronts between consecutive reconstruction loops. For situations in which the wavefront did not meet the convergence criterion, a maximum number of 20 reconstruction loops was specified to reduce simulation time in the case of poor convergence or non-convergence (e.g. when unsuitable plane locations were chosen).

Due to speed considerations, it is important to minimize the number of GS loops required for wavefront reconstruction by optimizing the value of the convergence criterion. In order to determine the optimal convergence requirement for our simulations, we tested values of 0.1 nm, 1 nm, 10 nm, 15 nm RMS. We found that using a convergence criterion of 1 nm RMS provided the best tradeoff between speed and accuracy. We therefore chose to use a 1 nm criterion for all subsequent atmospheric simulations that use a Kolmogorov power spectrum.

\subsection{Phase Unwrapping}\label{sec:phase_unwrapping}

A consequence of the numerical GS reconstruction method is that the phase can only be extracted in a range of $\pm \pi$. Wavefront aberrations larger than one wave peak-to-valley will result in a wrapped phase, causing abrupt discontinuities to appear in the reconstruction. This effect can inhibit wavefront correction in practice when using continuous deformable mirror surfaces. In addition, our metric for determining reconstruction accuracy --- the RMS of the residual WFE --- is affected by phase unwrapping, as multiple waves of phase wrapping would appear as a poor reconstruction in that metric. 

Many solutions have been proposed for 2D phase unwrapping, each with their own sets of strengths and weaknesses.\cite{schofield_03} After investigating several phase unwrapping methods, we chose the advanced ``\textit{lspv}'' algorithm from the WaveProp package\cite{Brennan:16,Brennan:17}. While the \textit{lspv} algorithm does not work well at sharp wavefront discontinuities, such as that of the pupil boundary, the localized errors at this boundary are consistently limited to the outer edges of the pupil. In order to alleviate this issue, the edges of the pupil were excluded from RMS residual WFE calculations; only the area within 95\% of the original pupil diameter was used for calculating the RMS residual WFE.

\section{Results and Discussion}\label{sec:results}

\subsection{Optimizing Detector Plane Locations}\label{sec:defocus_distance_analysis}

In order to determine the optimal measurement plane locations, we identified the spatial frequency scale of atmospheric aberrations as the most likely factor to impact the ideal propagation distances for a given aperture size and wavelength. As mentioned in the introduction, the Talbot effect describes how phase aberrations in one location lead to intensity variations at distances that depend on the spatial frequency of the aberration. To study the effects of the atmosphere on measurement plane distances for the nlCWFS, we used the ratio $D/r_0$ as a unitless metric that can be scaled for other systems to characterize turbulence strength.

The symmetric four-plane configuration was tested using a range of $D/r_0$ values (4, 8, 16, 32, and 64), which are representative levels of turbulence experienced by a WFS during closed-loop operation. All atmospheric simulations used a series of 16 randomly-seeded input wavefront aberrations that follow a Kolmogorov power spectrum, with a separate set being used for each $D/r_0$ value. This provides sufficient turbulence profiles to demonstrate representative performance of the sensor while also maintaining an acceptable simulation run-time. The simulations used a pupil diameter of $D=0.5$ m and monochromatic light at wavelength $\lambda = 532$ nm. The inner measurement planes (P1 and P2) were moved within a range from $\pm 1$ km to $\pm 120$ km, while the outer measurement planes (P3 and P4) were moved from $\pm 10$ km to $\pm 250$ km. These values were chosen so as to pass through the Talbot distance corresponding to the spatial size of the Fried parameter (i.e. setting $a=r_0$ in Equation \ref{eq_talbot_dist}), and were moved from the pupil to greater distances to study how diffraction affects sensing at greater distances. The resolution used to explore the distance parameter space was adjusted for each $D/r_0$ value to achieve acceptable computation run-time given the array size requirements (\S \ref{sec:img_gen}).

\begin{sidewaysfigure}
    \centering
    \includegraphics[width=\textwidth]{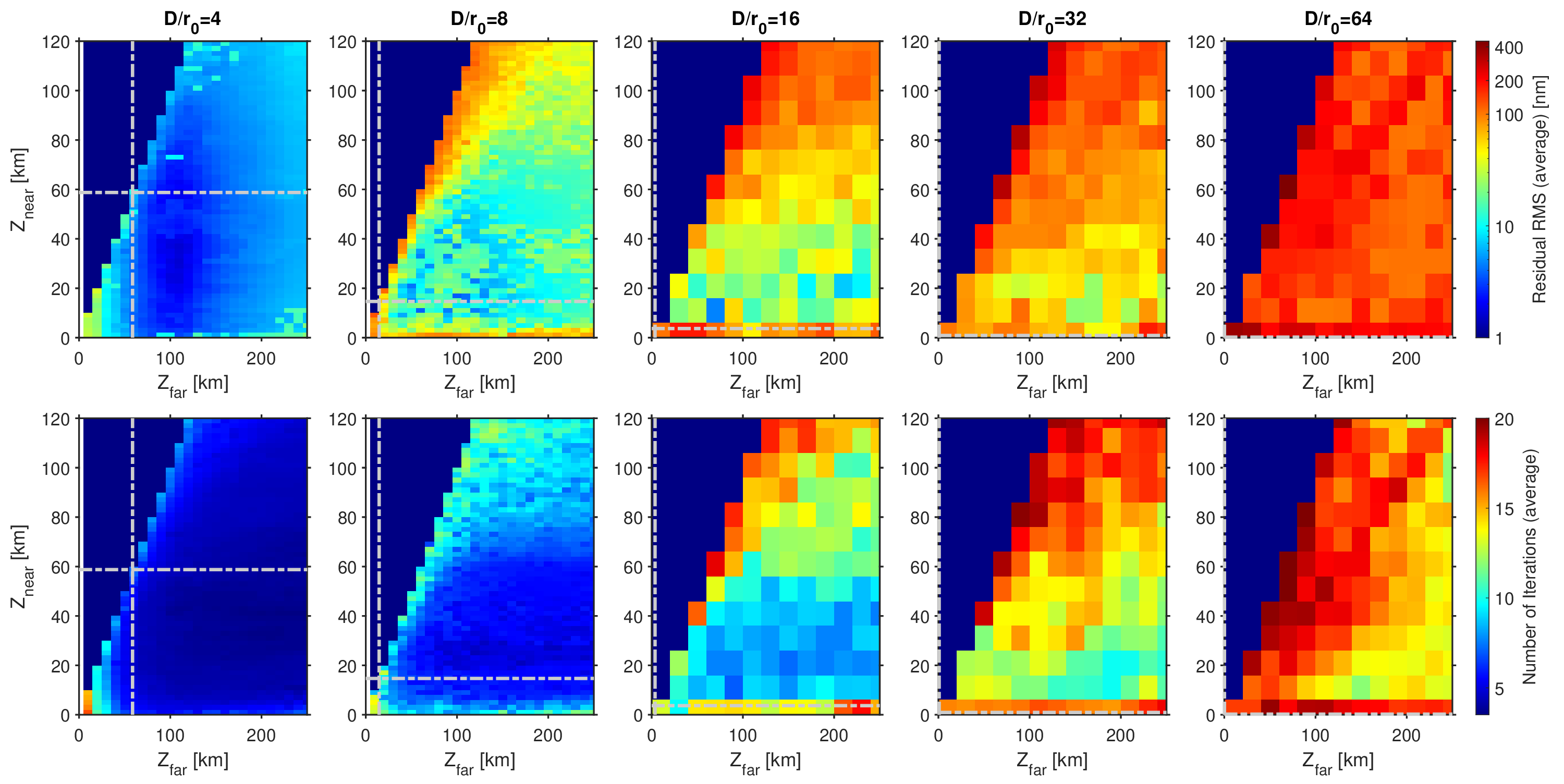}
    \caption{Symmetric four-plane reconstruction of Kolmogorov turbulence showing average values of the RMS of the residual WFE (top row) and the number of iterations to reach a converged solution (bottom row) for different values of $D/r_0$. The Talbot distance corresponding to each $r_0$ value are shown as horizontal gray lines for $Z_{\rm near}$ and as vertical gray lines for $Z_{\rm far}$. Note that the upper-left portion of the plots have no data as this region is where $Z_{\rm near}$ is larger than $Z_{\rm far}$. Since this parameter space is symmetrically identical to the region already plotted, this space was set to NaN values to reduce simulation time.}
    \label{fig:4_plane_kolmogorov_Dr0}
\end{sidewaysfigure}

Relevant performance metrics recorded for each combination of input parameters included: $\sigma_{\rm RMS}$, the RMS of the residual wavefront error (WFE) of the interior 95\% of the pupil diameter; $N_{\rm iter}$, the number of iterations needed for convergence; and the time to complete the reconstruction. These metrics were then averaged over all 16 simulated wavefronts to improve the statistical validity of the results. Results for this analysis are shown in Figure \ref{fig:4_plane_kolmogorov_Dr0}. 

We find that poor reconstructions (large values of $\sigma_{\rm RMS}$ and $N_{\rm iter}$) occur when either $Z_{\rm near}$ or $Z_{\rm far}$ are too close to the pupil. In this regime, the poor performance is caused by a lack of substantial diffraction, providing the sensor too little information about the pupil phase. Poor performance also occurs when the measurement planes are too close to one another due to an effective repeat of information, i.e. the planes are so similar that the sensor is essentially acting as a two-plane configuration. These results hold for all simulated $D/r_0$ values.

Optimal reconstructions (low $\sigma_{\rm RMS}$ and $N_{\rm iter}$) tend to occur when the inner planes are located beyond the Talbot distance, corresponding to an aberration with a spatial period of $a=r_0$, i.e. where:

\begin{equation}
Z_{\rm near} = \frac{2 a^2}{\lambda} \approx \frac{2 r_0^2}{\lambda}.
\end{equation}

\noindent This location is indicated by a horizontal gray line in Figure~\ref{fig:4_plane_kolmogorov_Dr0} and can be seen to change depending on $D/r_0$.

The optimal $Z_{\rm far}$ location did not easily map to a specific value or multiple of the Talbot distance. This is indicated by the lack of correlation between $Z_{\rm far}$ distances that offer quality reconstructions and the $r_0$ Talbot distance, denoted by a vertical gray line in Figure~\ref{fig:4_plane_kolmogorov_Dr0}. Instead, we find that there exists a minimum $Z_{\rm far}$ value that offers high-quality reconstructions provided that $Z_{\rm far}$ exceeds a certain (threshold) distance. As shown in Figure~\ref{fig:4_plane_kolmogorov_Dr0}, there exists a broad parameter space for each $D/r_0$ simulation where the symmetric, four-plane nlCWFS offers low values for both $\sigma_{\rm RMS}$ and $N_{\rm iter}$, provided that $Z_{\rm far}$ is larger than approximately $50$ km. 

\begin{figure}
    \centering
    \includegraphics[width=\textwidth]{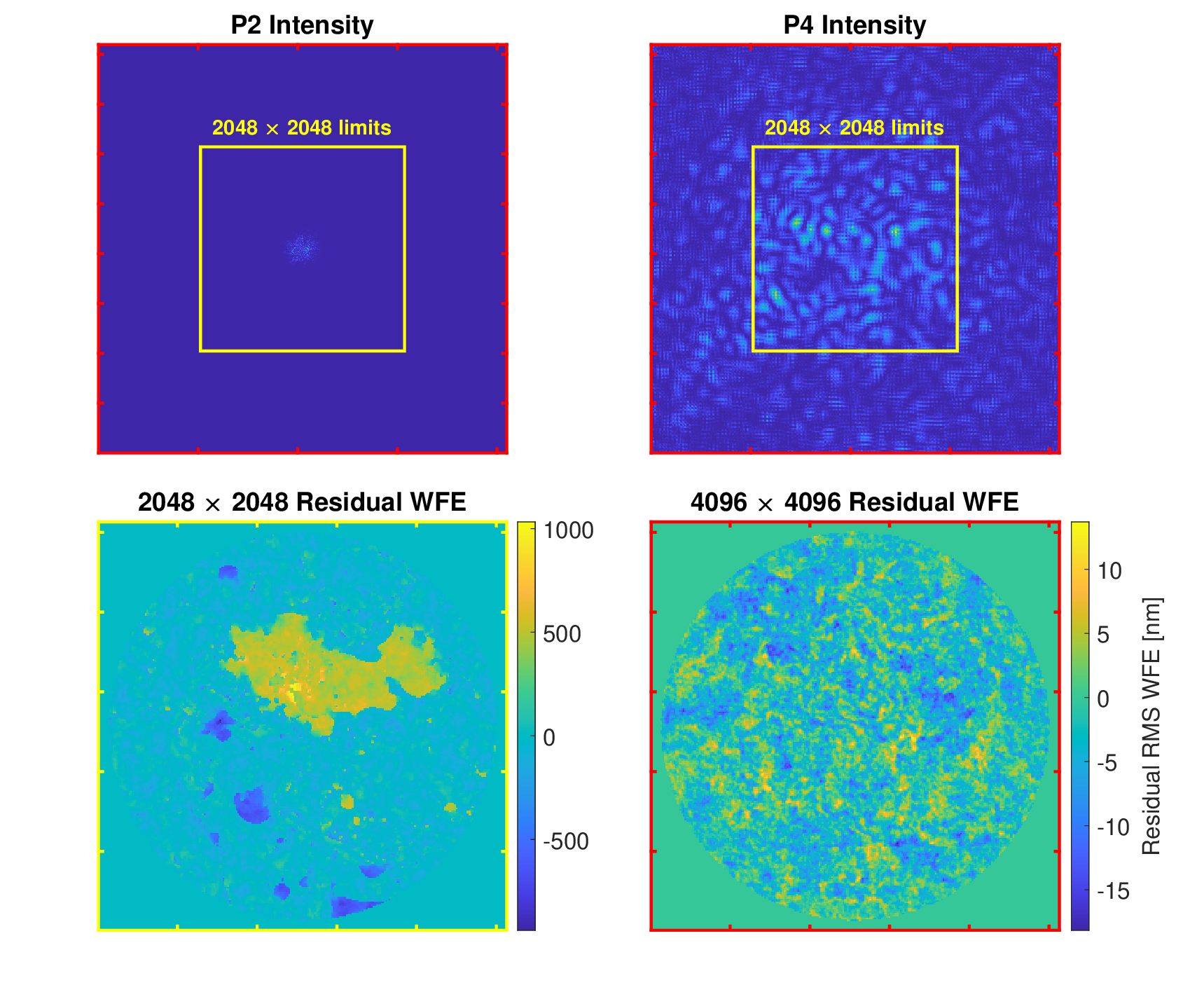}
    \caption{Comparison of the difference in the P2 and P4 detector planes (top) and reconstruction quality (residual RMS WFE, bottom) for $Z_{\rm near} = 9$ km and $Z_{\rm far} = 190$ km when the detector array size is changed from $2048{\times}2048$ (yellow) to $4096 {\times}4096$ (red).}
    \label{fig:Dr0_32_array_clipping}
\end{figure}

\begin{figure}
    \centering
    \includegraphics[width=\textwidth]{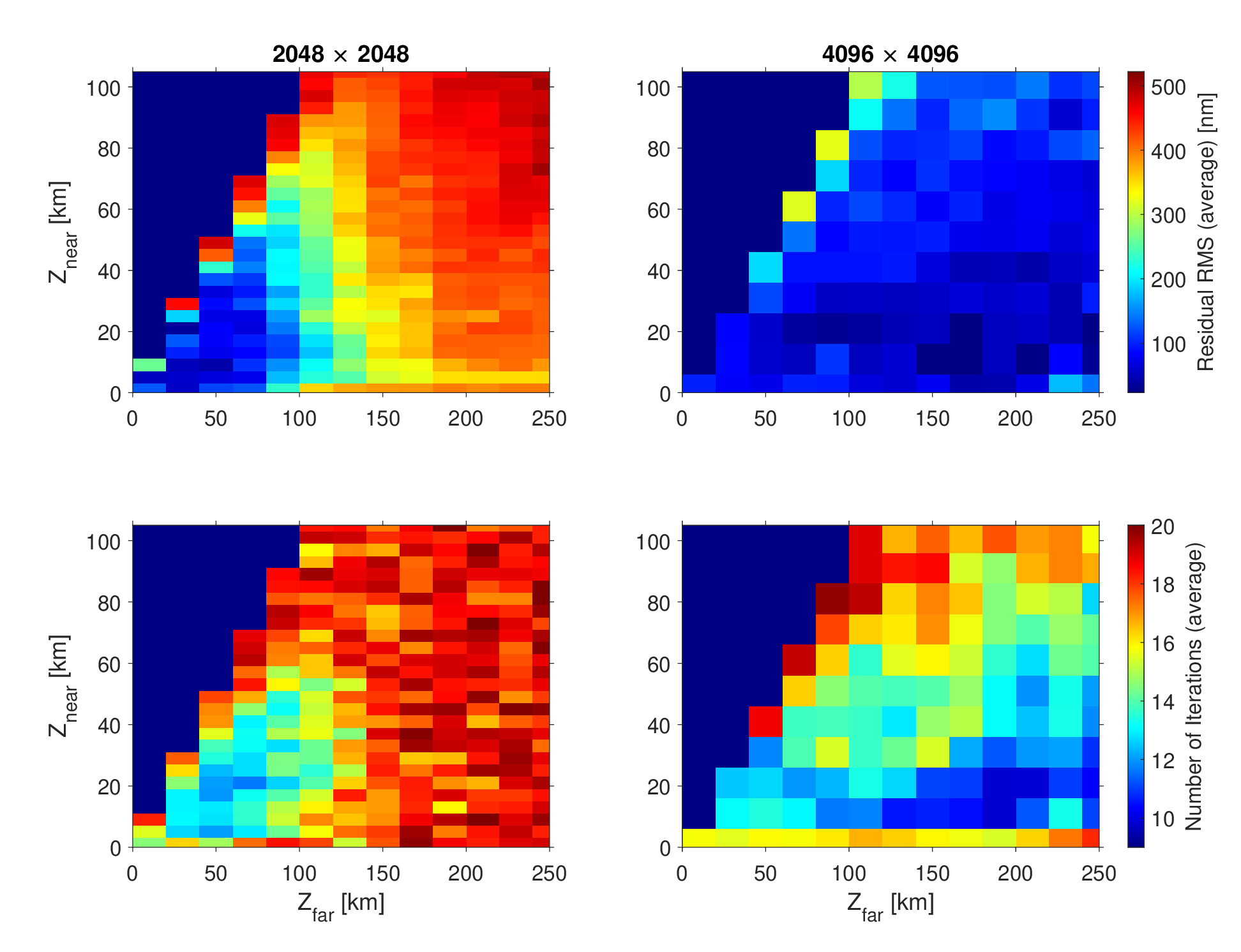}
    \caption{Comparison of two simulations of $D/r_0 = 32$, identical to that in Figure \ref{fig:4_plane_kolmogorov_Dr0}, but with one using $2048{\times}2048$ sized arrays for the measurement planes, and the other using $4096{\times}4096$ sized arrays. The effect of beam clipping on reconstruction accuracy can be seen in the smaller array size.}
    \label{fig:Dr0_32_comparison}
\end{figure}

Using our simulation results to calibrate the optimal $Z_{\rm far}$ locations, we find, through conservation of Fresnel number ($F=D^2 /Z\lambda$), that:
\begin{equation}
Z_{\rm far} \geq C D^2 / \lambda,
\label{eq:z_far_param}
\end{equation}
where $\lambda$ and $D$ are in units of meters and $C \approx 0.106$ is a coefficient that has been determined empirically. Optimal performance is achieved once the outer planes are located beyond this $Z_{\rm far}$ value, so long as $Z_{\rm near}$ is positioned appropriately. In practice, the outer planes cannot be moved further away from the pupil without limit because the finite size of detectors will eventually start to lose intensity information via beam clipping. This is particularly the case for stronger turbulence ($D/r_0$ values of 16, 32, 64) where a failure to fully capture the diffracted beam can lead to reconstruction errors. In initial simulations, the detector array size for these $D/r_0$ values was limited to $2048{\times}2048$. However, with the turbulence strength leading to increased diffractive effects during propagation, at the outer planes (P3 and P4, which are governed by $Z_{\rm far}$), this scaling did not fully capture the simulated light in the outer planes as shown in Figure~\ref{fig:Dr0_32_array_clipping}. Providing the nlCWFS reconstruction algorithm with such incomplete information led to inaccuracies in the reconstructed wavefront. This is shown in Figures~\ref{fig:Dr0_32_array_clipping} and \ref{fig:Dr0_32_comparison} where the undersampled $2048{\times}2048$ array size is compared to the adequately sampled $4096{\times}4096$ case which shows improved performance. These results indicate that the value $Z_{\rm far}$ can only be increased up to the distance at which the outer planes cease to fully capture the diffracted light, and this is why sampling of $4096{\times}4096$ has been used in the studies of higher $D/r_0$ values.

We tested and confirmed that, by conserving Fresnel number, the minimum $Z_{\rm far}$ distance results could be replicated for any size pupil diameter and observing wavelength. Our simulations show that changing $D$ and/or $\lambda$ produces nearly identical results as those shown in Figure~\ref{fig:4_plane_kolmogorov_Dr0}, provided that the Fresnel number is conserved. In other words, the minimum $Z_{\rm far}$ may be calculated based on scaling relations for a different optical system or telescope using the Fresnel number, as done in Equation \ref{eq:z_far_param}. Various examples of converted values for different $D$ and $\lambda$ values are shown in Table \ref{tab:conversion}.

The combination of these results demonstrates that the optimal defocus distances for the nlCWFS encompass a wide range of plausible values, which is beneficial for the prospect of practical applications. In summary, the symmetric, four-plane nlCWFS appears to show optimal or near-optimal performance (low residual WFE, and low number of iterations) as long as the following two factors are met:
\begin{enumerate}
    \item The inner ($Z_{\rm near}$) planes should be located further than the Talbot distance corresponding to an aberration with a spatial period of $r_0$.
    \item The outer ($Z_{\rm far}$) planes depend on the system Fresnel number and should be located beyond $Z_{\rm far} \geq C D^2 / \lambda$ where $C \approx 0.106$, yet closer than the distance at which the diffracted light can no longer be fully captured by detector arrays.
\end{enumerate}

These results suggest that there is flexibility when choosing measurement plane distances and indicate that the nlCWFS is very adaptable and resilient to ``incorrect'' placement of its measurement planes. When designing a physical implementation of the nlCWFS, consideration should be give to the expected range of effective $D/r_0$ values which will be experienced by the WFS to ensure optimum inner plane positioning under various conditions. Additionally, while this work has assumed monochromatic illumination, given the resilience of the nlCWFS to plane placement and previous work regarding broadband illumination of the nlCWFS, we expect similar trends would be noted in the polychromatic use case \cite{letchev_22}.

\begin{table}[]
\caption{Measurement plane positions ($Z_{\rm near}$ and $Z_{\rm far}$ values) for various values of beam diameter and wavelength given a fixed value of $D/r_0=8$. Values are determined by conserving Fresnel number. These represent the conversion from full telescope and simulated nlCWFS (as presented in this study) to a physical implementation of the sensor.}
\label{tab:conversion}
\resizebox{\textwidth}{!}{%
\begin{tabular}{|l|llllllll|c|}
\hline
\multicolumn{1}{|c|}{\multirow{2}{*}{$Z$ {[}m{]}}} & \multicolumn{8}{c|}{Beam Diameter ($D$),   with $D/r_0 = 8$}                                                                                                                                                                & \multirow{2}{*}{$\lambda$ {[}microns{]}} \\
\multicolumn{1}{|c|}{}                           & \multicolumn{1}{c}{5 mm} & \multicolumn{1}{c}{2 cm} & \multicolumn{1}{c}{25 cm} & \multicolumn{1}{c}{0.5 m} & \multicolumn{1}{c}{2 m} & \multicolumn{1}{c}{4 m} & \multicolumn{1}{c}{8 m} & \multicolumn{1}{c|}{30 m} &                                                       \\ \hline
near                                             & 1.50 $\times 10^{-2}$    & 24.0                       & 3.75 $\times 10^3$        & 1.50 $\times 10^4$        & 2.40 $\times 10^5$      & 9.60 $\times 10^5$      & 3.84 $\times 10^6$      & 5.40 $\times 10^7$        & \multirow{2}{*}{0.532}                                \\
far                                              & 5.00 $\times 10^{-2}$    & 80.0                       & 1.25 $\times 10^4$        & 5.00 $\times 10^4$        & 8.00 $\times 10^5$      & 3.20 $\times 10^6$      & 1.28 $\times 10^7$      & 1.80 $\times 10^8$        &                                                       \\ \hline
near                                             & 7.79 $\times 10^{-3}$    & 12.5                     & 1.95 $\times 10^3$        & 7.79 $\times 10^3$        & 1.25 $\times 10^5$      & 4.99 $\times 10^5$      & 2.00 $\times 10^6$      & 2.81 $\times 10^7$        & \multirow{2}{*}{1.024}                                \\
far                                              & 2.60 $\times 10^{-2}$    & 41.6                     & 6.49 $\times 10^3$        & 2.60 $\times 10^4$        & 4.16 $\times 10^5$      & 1.66 $\times 10^6$      & 6.65 $\times 10^6$      & 9.35 $\times 10^7$        &                                                       \\ \hline
near                                             & 5.15 $\times 10^{-3}$    & 8.24                     & 1.29 $\times 10^3$        & 5.15 $\times 10^3$        & 8.24 $\times 10^4$      & 3.29 $\times 10^5$      & 1.32 $\times 10^6$      & 1.85 $\times 10^7$        & \multirow{2}{*}{1.55}                                 \\
far                                              & 1.72 $\times 10^{-2}$    & 27.5                     & 4.29 $\times 10^3$        & 1.72 $\times 10^4$        & 2.75 $\times 10^5$      & 1.10 $\times 10^6$      & 4.39 $\times 10^6$      & 6.18 $\times 10^7$        &                                                       \\ \hline
\end{tabular}%
}
\end{table}

\subsection{Determining the Optimal Number of Planes}\label{sec:plane_num_analysis}

Changing the number of planes can potentially impact the reconstruction process in terms of both accuracy and speed. For example, fewer measurement planes reduce the number of Fourier transforms needed per reconstruction loop and thus may allow for faster reconstruction times. In addition, fewer planes may allow for a more compact optical design as well as increased flux per plane, as fewer splitting optics are required. In contrast, more measurement planes may allow for more efficient algorithmic convergence, allowing the reconstructor to use fewer iterations and potentially increase its speed. In an effort to search for any benefits in changing the number of planes, we studied nlCWFS designs that use three planes and five planes to compare to the previous analysis of the symmetric-four-plane design.

\subsubsection{Three-Plane Design}\label{sec:kolmogorov_threeplane_results}

\begin{sidewaysfigure}
    \centering
    \includegraphics[width=\textwidth]{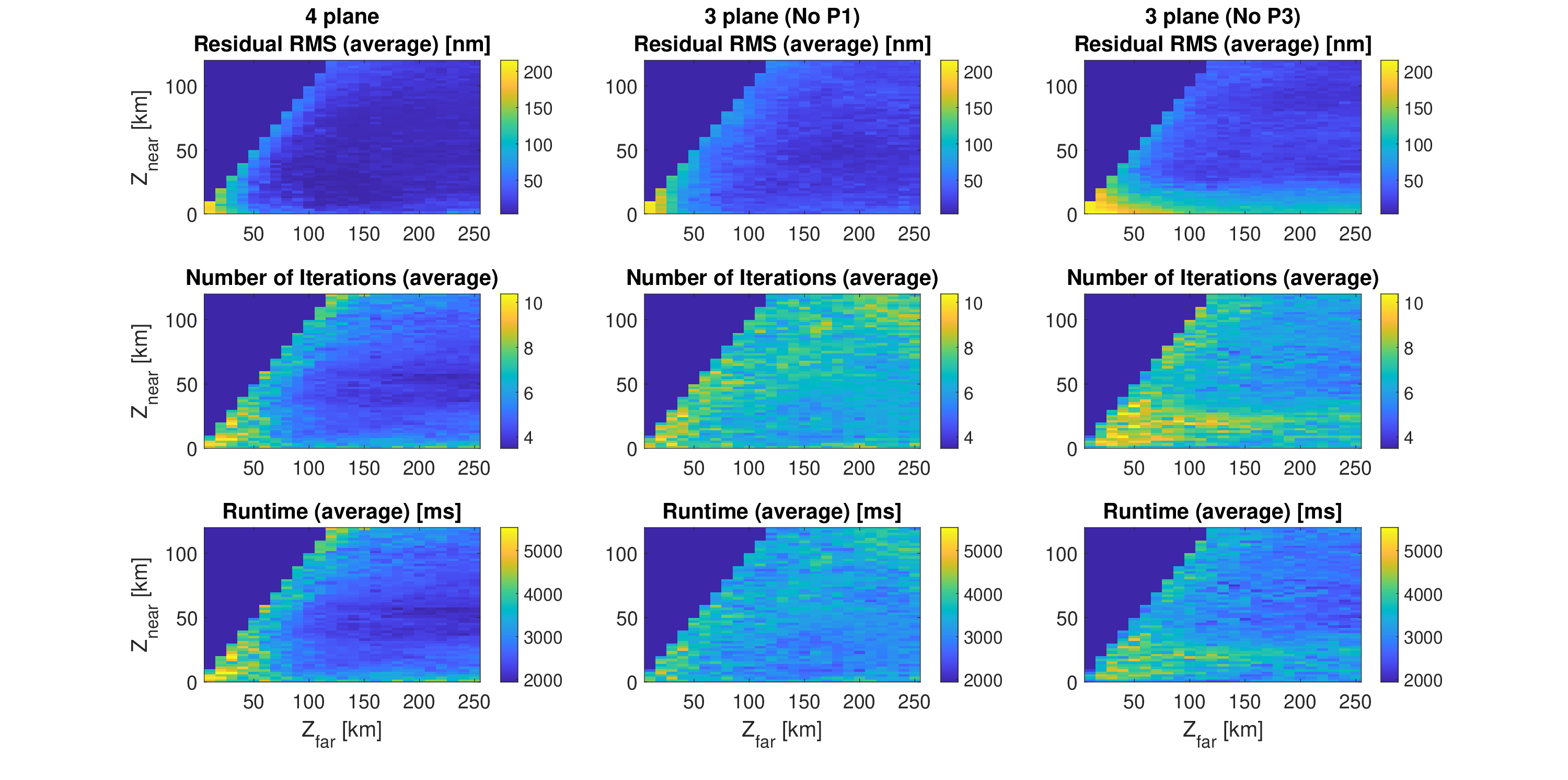}
    \caption{Three-plane and four-plane reconstructions of a Kolmogorov phase aberration with a $D/r_0=4$, showing the average values of the RMS WFE (top row), the average number of iterations (middle row), and the approximate average reconstruction time based on the average time per loop (bottom row). The reconstruction times are not based on optimized algorithms and are only intended as a relative comparison.}
    \label{fig:3_plane_kolmogorov}
\end{sidewaysfigure}

The three-plane configuration was tested using the same layout and simulation design as the symmetric-four-plane configuration, except one plane was removed, and only $D/r_0=4$ was evaluated. For one set of simulations, the P3 ($-Z_{\rm far}$) plane was removed, and for the second set of simulations, the P1 ($-Z_{\rm near}$) plane was removed.

The results for these two sets of simulations are shown in Figure \ref{fig:3_plane_kolmogorov}. Both three-plane configurations produce RMS residual WFE values comparable to the symmetric four-plane configuration ($\sigma_{\rm RMS} \approx 10$ nm). However, both sets of simulations require an increased number of iterations (6-8 compared to around 4) to reach the same RMS WFE. Despite a shorter time per iteration (a three-plane loop took approximately 88\% of the time of a four-plane loop including overheads), the increased number of iterations compared to the four-plane configuration resulted in an increased overall latency. Therefore, our simulations indicate that, unless the pragmatics of hardware implementation (e.g. optical design, illumination levels, noise performance, detector size etc.) outweigh speed demands, the three-plane configuration does not offer an improvement over the four-plane design.

\subsubsection{Five-Plane Design}\label{sec:kolmogorov_fiveplane_results}

\begin{figure}
    \centering
    \includegraphics[width=\textwidth]{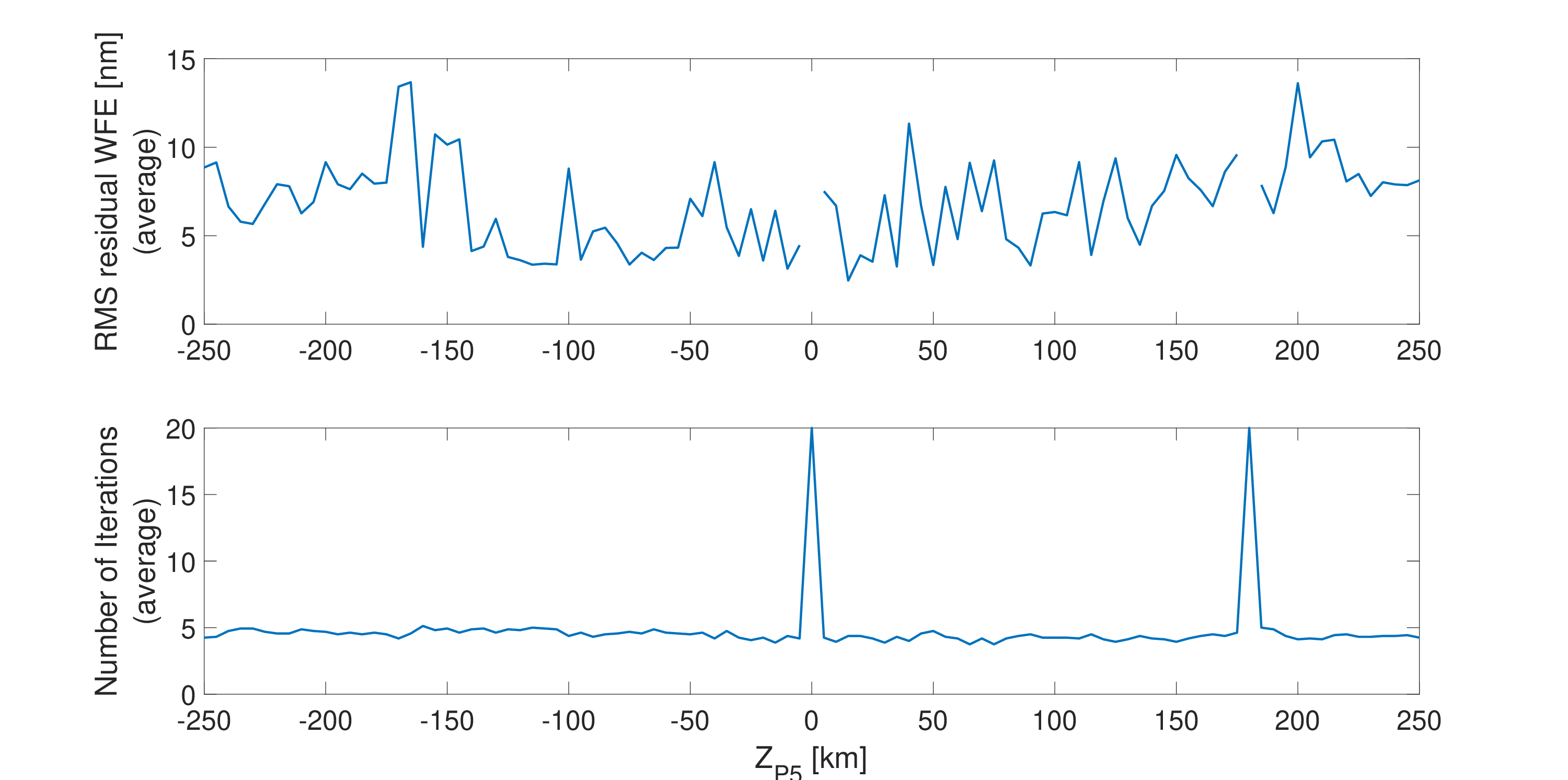}
    \caption{Five-plane reconstructions of a Kolmogorov phase aberration with $D/r_0 = 4$, showing average values of the RMS WFE (top) and the average number of loop iterations required to reach a $\sigma_{\rm RMS}=1$ nm convergence criterion (bottom). Gaps in the RMS plot and peaks in the iteration plot correspond to zero-distance propagations (i.e. when P5 was placed at the location of either P4 or the pupil). Since the P5 plane was placed in between P4 and the pupil in the modified GS loop, these locations would result in the algorithm attempting to propagate by a zero distance, and results in numerical errors that inflate the RMS WFE and number of iterations.}
    \label{fig:5_plane_kolmogorov}
\end{figure}

The results of the three-plane design suggested that adding additional measurement planes may provide a reconstruction time advantage through faster convergence. Therefore, a design with five planes was studied. To facilitate an efficient study, we chose the best $Z_{\rm near}$ and $Z_{\rm far}$ values from the four-plane results and added a single extra plane, varying its distance through the same range of values as before (Z = -250 to 250 km). The set of parameters used are $Z_{\rm near}=47$ km and $Z_{\rm far}=180$ km with a $D/r_0=4$. Results are shown in Figure \ref{fig:5_plane_kolmogorov}. 

We find that the average RMS residual WFE varies between $\sigma_{\rm RMS} \approx 3 - 14$ nm, and average number of iterations typically range between $N_{\rm iter} = 4 - 5$. For reference, the symmetric four-plane values at those same distances were $\sigma_{\rm RMS} \approx 6.5$ nm and $N_{\rm iter}=4.2$ iterations, respectively. This result indicates that the five-plane configuration offers limited improvement in RMS WFE compared to the four-plane design, while also not offering any improvement in iteration number, causing the overall latency to increase. This combination of factors indicates that increasing the number of nlCWFS measurement planes beyond four offers minimal benefit.

\subsection{Verifying Robustness of Symmetric Four-Plane Configuration using Monte Carlo Simulations}\label{sec:monte_carlo_test}

In order to verify that other, potentially asymmetric, four-plane configurations would not substantially improve results compared to the nominal symmetric configuration, we performed Monte Carlo simulations that chose four random propagation distances for the measurement plane locations. Using the same parameters as in previous simulations (wavelength, telescope diameter, etc.), the four distances were limited to $\pm1000$ km from the pupil. To maintain the same naming convention and GS-loop propagation order as before (e.g. consistent with Figure \ref{fig:plane_config}), each set of four plane locations was sorted so that they were in the order: P3, P1, P2, and P4, from the most negative to the most positive distance along the optical propagation axis. Simulated results for the intensity at each measurement plane were then passed through the same reconstruction pipeline as before.

To explore the propagation-distance parameter space, a series of 4,000 unique four-plane distance combinations were generated. Each combination was studied using 16 aberration maps with $D/r_0=4$. The results were then sorted by $\sigma_{\rm RMS}$ and compared with the four-plane symmetric design. All other parameters and metrics of interest, such as $N_{\rm iter}$ and individual propagation distances, were sorted alongside the WFE to study any patterns that could emerge.

\begin{figure}
    \centering
    \includegraphics[width=\textwidth]{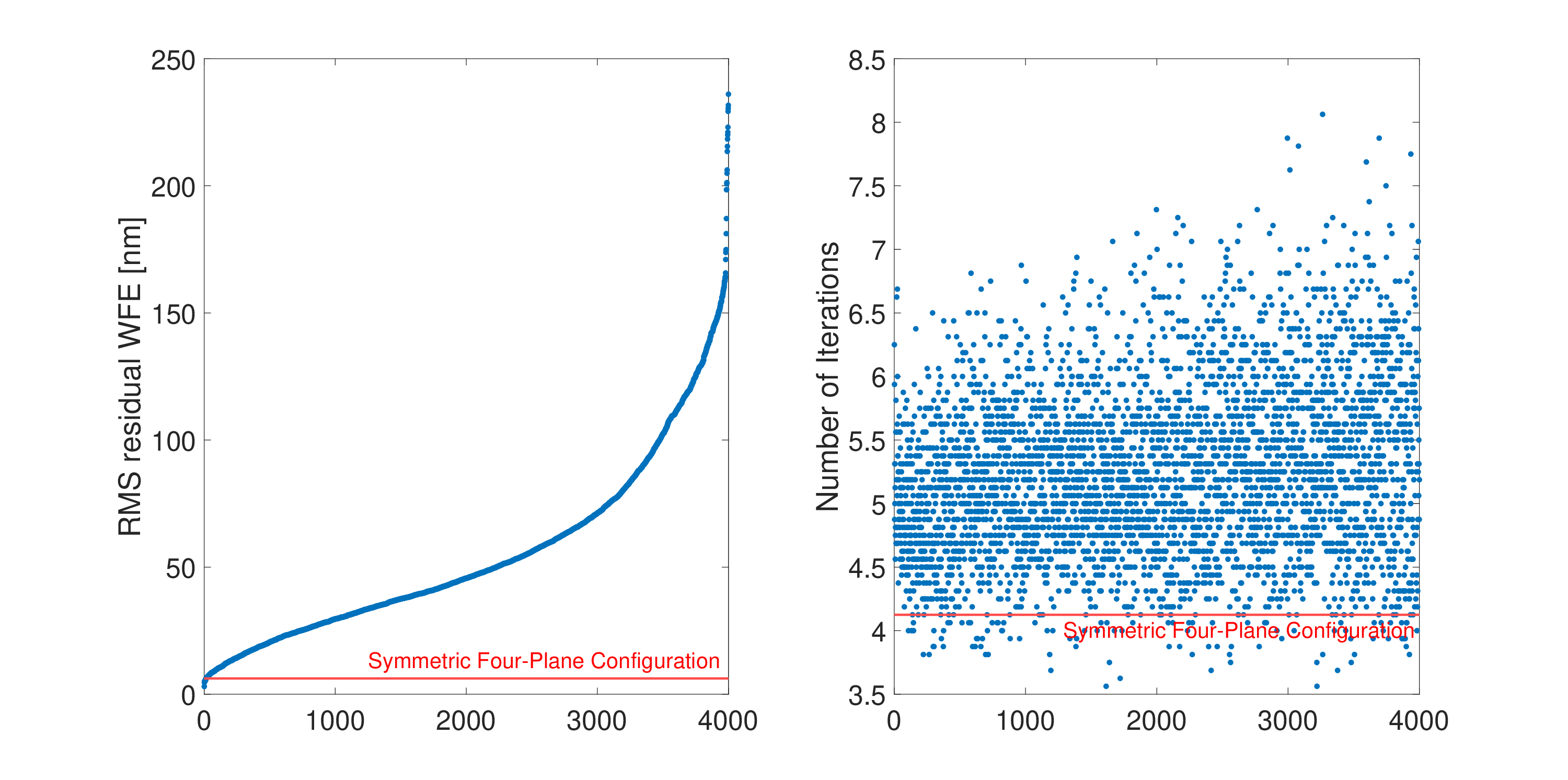}
    \caption{The RMS WFE and number of iterations required for convergence for 4000 randomly selected sets of propagation distances for a nlCWFS with four measurement planes. Both RMS WFE and iterations were averaged for 16 different aberrations with $D/r_0 = 4$. The ``close to best'' symmetric-four-plane configuration values for each quantity are shown as horizontal red lines. There are no randomly selected distance values that produce both an RMS residual WFE and number of iterations that are smaller than the symmetric-four-plane values.}
    \label{fig:MC_RMS_iter_plot}
\end{figure}

Figure~\ref{fig:MC_RMS_iter_plot} shows results for the Monte Carlo random propagation distance simulations in terms of RMS WFE and number of reconstructor loop iterations. We find that fewer than 0.4\% of the random distance combinations produce an RMS WFE lower than the symmetric four plane design, and 0\% produce both a lower $\sigma_{\rm RMS}$ and smaller value of $N_{\rm iter}$ than the symmetric four plane configuration. 

\begin{figure}
    \centering
    \includegraphics[width=0.5\textwidth]{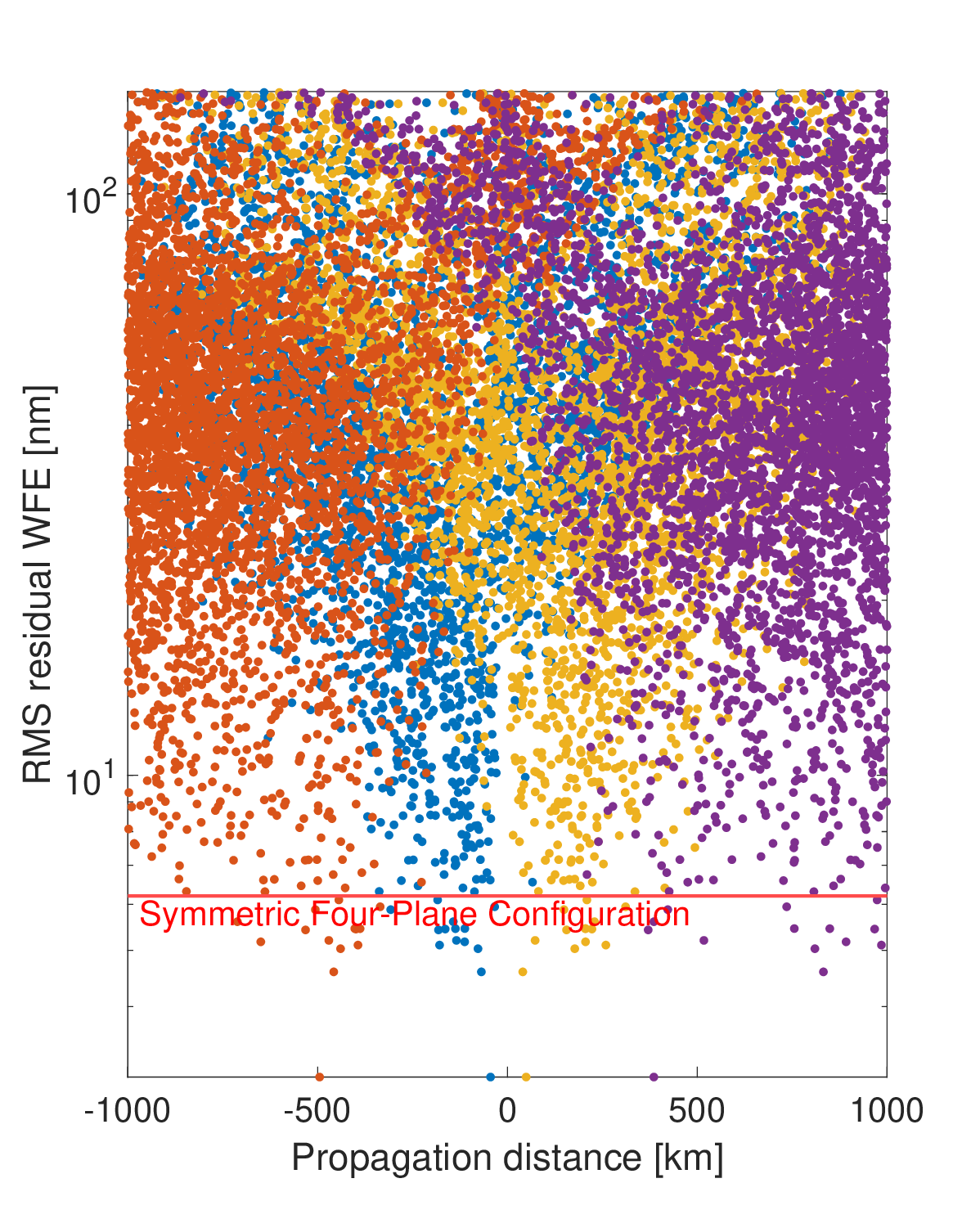}
    \caption{The locations of four measurement planes with randomly selected propagation distances, sorted by RMS residual WFE, averaged for 16 different aberrations with $D/r_0 = 4$. The locations of each plane are shown in different color (P1: blue, P2: yellow, P3: purple, P4: orange). The ``close to best'' symmetric-four-plane configuration value of RMS residual WFE is shown as a horizontal red line.}
    \label{fig:MC_z_dist_RMS_plot}
\end{figure}

Another way of interpreting the Monte Carlo simulations is shown in Figure \ref{fig:MC_z_dist_RMS_plot}. A pattern in the random propagation distance values can be noted when the results are sorted by WFE. The best performance occurs when two measurement planes are placed on either side of the optical system pupil, and, as the plane locations become more asymmetric, the WFE increases. In fact, most of the distance combinations that result in WFEs smaller than the symmetric four-plane configuration contain two planes on either side of the pupil, with two being closer to the pupil and two being further away. This lends strong support to the hypothesis that the symmetric four-plane configuration is the optimal configuration for the nlCWFS.

\begin{figure}
    \centering
    \includegraphics[width=\textwidth]{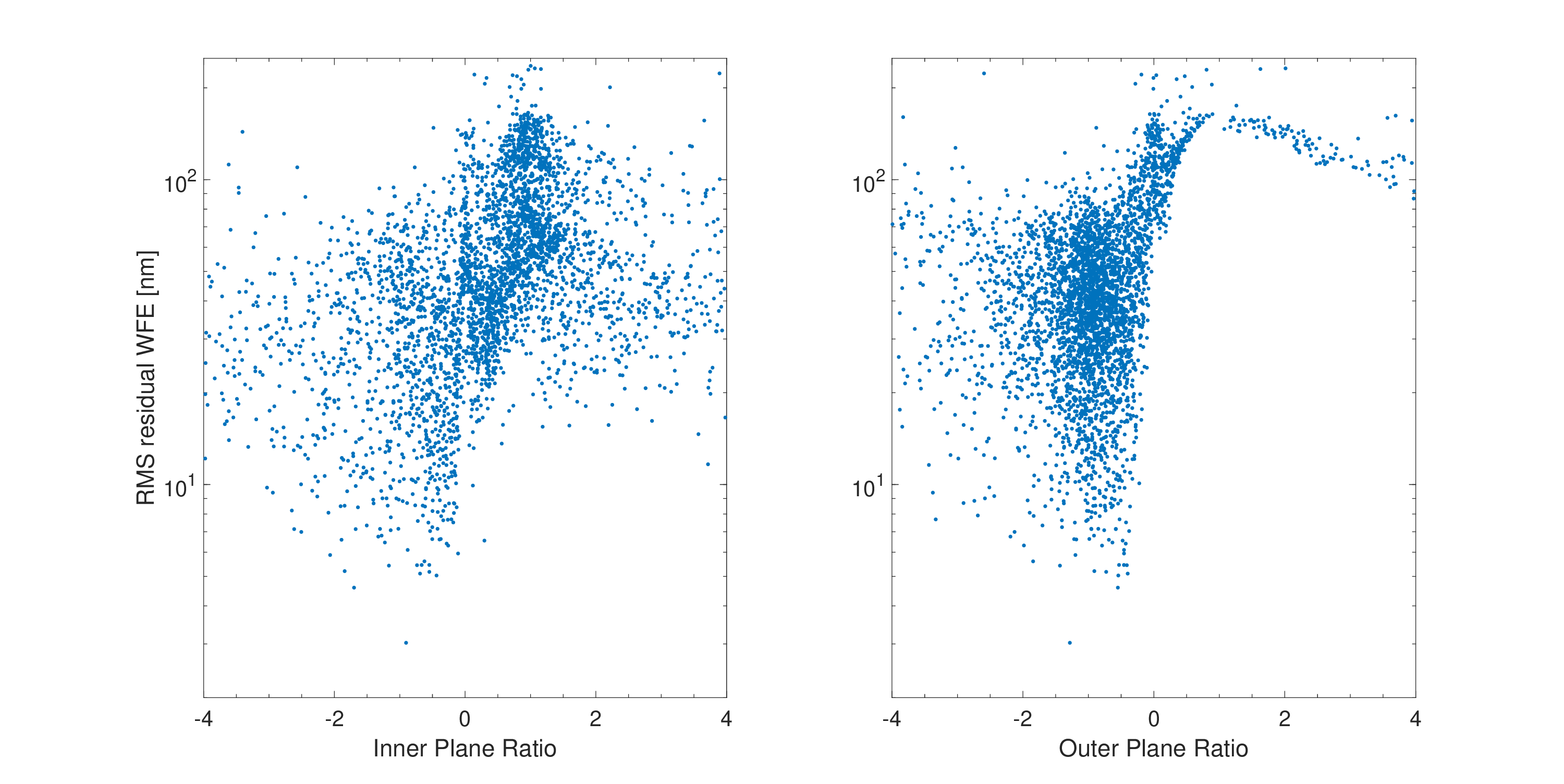}
    \caption{The ratios of the two inner and outer measurement plane distances, respectively for 4000 randomly selected sets of propagation distances, sorted by RMS residual WFE, averaged for 16 different aberrations with $D/r_0 = 4$. Negative values for the ratio indicate the measurement planes are on opposite sides of the system pupil, while positive values indicate the planes are on the same side.}
    \label{fig:MC_RMS_ratio_plot}
\end{figure}

To study how the level of symmetry between measurement planes impacts reconstruction accuracy, Figure \ref{fig:MC_RMS_ratio_plot} plots WFE as a function of the ratio of inner plane distances ($Z_1/Z_2$) and also the ratio of outer plane distances ($Z_3/Z_4$). We find that the best results occur when $Z_1/Z_2$ and $Z_3/Z_4$ are within a factor of two of each other (a ratio of 0.5-2), indicating that the planes are approaching being equidistant and located on opposite sides of the pupil. These results corroborate the idea that symmetry is important for wavefront reconstruction quality. As reconstruction accuracy degrades, the envelope of Monte Carlo results for both the inner plane ratio and outer plane ratio drifts further from a symmetric configuration. This effect is most clearly evident in the outer planes, which are responsible for sensing low spatial frequency content and thus most of the power in atmospheric turbulence, where no positive $Z_3/Z_4$ combinations generated an RMS WFE smaller than 50 nm. Considering the results as a whole, both a systematic exploration of plane locations and a Monte Carlo simulation suggest that the symmetric four-plane configuration is likely an optimal choice for the nlCWFS design.

\subsection{Spatial Sampling Requirements}\label{sec:sampling}

The number of pixels used to sense light across the beam impacts the wavefront reconstruction process in terms of accuracy, sensitivity, and speed. In the previous simulations, spatial sampling was set to sufficiently high values to ensure that numerical noise did not influence results for identifying the best plane locations. In practice, the requirements on spatial sampling for the nlCWFS are more complicated than conventional sensors, as the planes of observation are not located at a pupil or focal plane, which are where most optical and turbulence parameters are defined.

\begin{figure}
    \centering
    \includegraphics[width=0.75\textwidth]{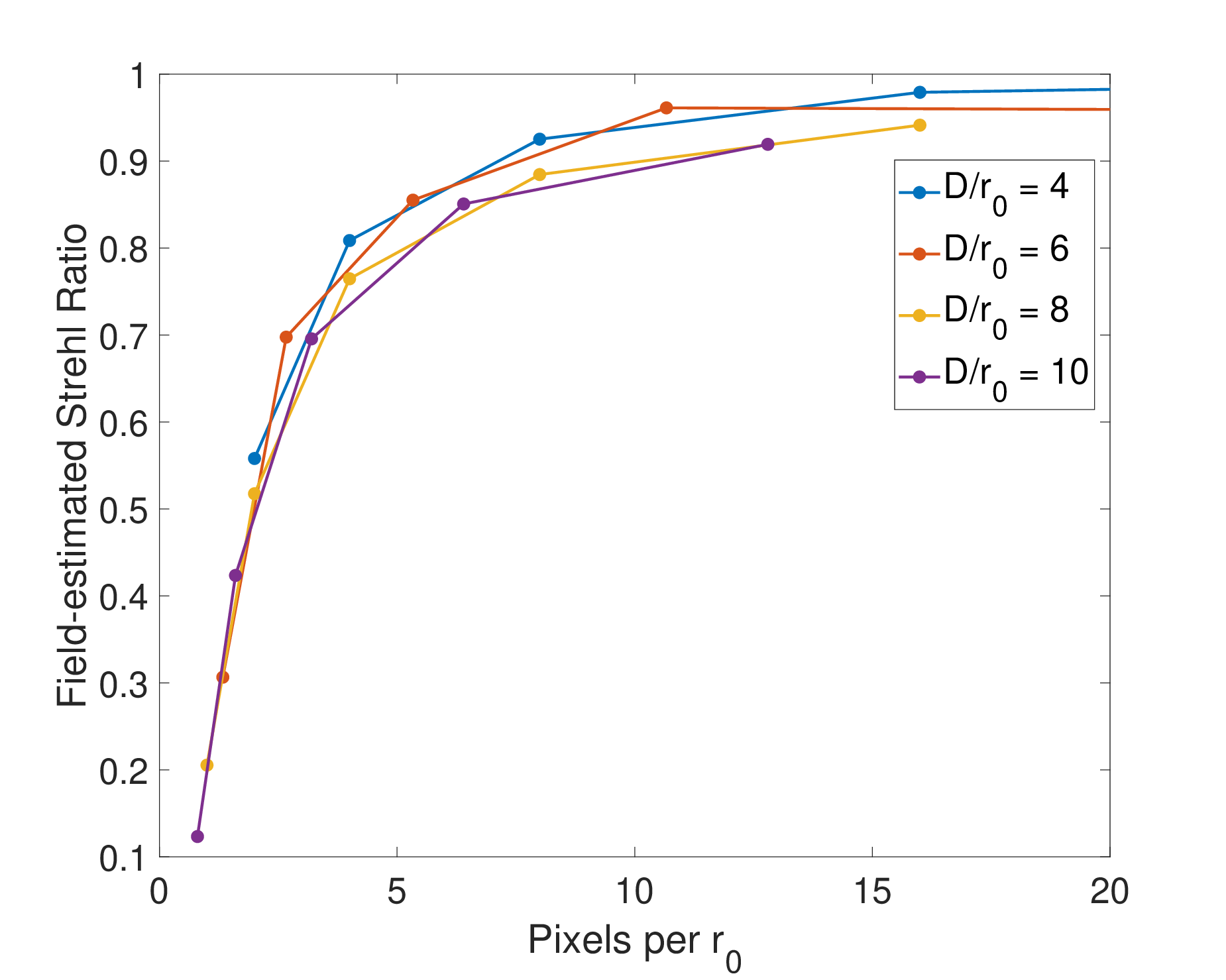}
    \caption{The impact of spatial sampling on field-estimated Strehl ratio as a function of turbulence strength. To achieve diffraction-limited performance, a Strehl ratio of at least 0.8 is required.}
    \label{fig:sampling_plot}
\end{figure}

To study how spatial sampling impacts the reconstruction process, we systematically adjusted the number of pixels relative to the value of $r_0$ across the system pupil. The simulations followed a similar procedure as in previous sections in that a series of 16 unique wavefront instances were generated for $D/r_0$ values of 4, 6, 8, and 10, after which the results were averaged. The wavefronts were then reconstructed using a symmetric four-plane nlCWFS with measurement planes located at the optimal propagation distances described in Section \ref{sec:defocus_distance_analysis}. An initial sampling rate of 128 pixels across the pupil diameter was used for both the input pupil wavefront. This increase in pupil sampling versus the 64 pixels used in previous studies ensured sufficient sampling for the $D/r_0 = 10$ case, and allowed better resolution in binning steps as part of this study. The pupil was zero-padded to a total array size of $4096{\times}4096$ for initial image generation, of which the inner $2048{\times}2048$ pixels were used for reconstruction. The resulting images were then binned for sampling tests by increasing factors of 2, 4, 8, and 16. Field-estimated Strehl ratio ($S$) was used as the primary metric for assessing performance\cite{banet_17}. Results of the analysis are shown in Figure \ref{fig:sampling_plot}. 

We find that the nlCWFS must sample the beam by at minimum $\approx 4 - 5$ pixels per $r_0$, as defined in the pupil plane, for a reconstruction that produces a field-estimated Strehl ratio sufficient for diffraction-limited performance ($S \approx 0.8$). This value is similar to results found for the Digital Holographic WFS \cite{banet_17}, which is not surprising given that both sensors rely on optical interference to reconstruct the wavefront.

\section{Conclusions}\label{sec:conclusions}

To achieve the performance requirements of the most demanding AO applications (astronomy, medical imaging, remote sensing, laser communications, etc.), WFSs that offer high sensitivity, a large dynamic range, and low latency are needed. Due to the physics-based trade-offs between these variables, the linearity of the wavefront retrieval process may need to be sacrificed. It is not unreasonable to expect that, as computational techniques and raw computing power continue to improve, the impact of nonlinearity on speed may be overcome.

The nlCWFS is a promising sensor for next-generation AO systems that is both sensitive and offers a large dynamic range. Yet, questions remain regarding its implementation in  both theory and practice. In particular, the number of measurement planes, their locations, and requisite spatial sampling have not been thoroughly explored.

In this paper, we performed numerical simulations to address these questions by quantifying WFE and studying reconstruction algorithm convergence while varying turbulence conditions and sensor parameters. We find that the ideal number of measurement planes for the nlCWFS is four, with two pairs of planes placed symmetrically on either side of the optical system pupil. This configuration is indeed the original design proposed by Guyon 2010\cite{guyon_10}. Although asymmetric designs produced successful reconstructions, they did not offer significantly improved accuracy, algorithm convergence, or speed. The three-plane, five-plane, and other randomly-selected permutations consistently demonstrate that the symmetric four-plane design outperformed all alternatives in one, if not all, of the performance metrics used.

Assuming Kolmogorov turbulence, for optimal performance in the symmetric four-plane design, the $Z_{\rm near}$ measurement planes should be located further than the Talbot distance corresponding to a spatial period of $a=r_0$, and the $Z_{\rm far}$ measurement planes should be located beyond $Z_{\rm far} \geq C D^2 / \lambda$ where $C \approx 0.106$. Adequate sensing was consistently achieved in such configurations provided that spatial sampling exceeds 4 - 5 pixels per $r_0$ as defined in the pupil plane for each of the four detectors, while also ensuring the full diffracted intensity is captured in all detector planes. When physically implementing a nlCWFS, the range of $D/r_0$ values experience by the WFS should be considered when determining the optimum plane locations. 

Future implementations of the nlCWFS will likely focus on improving its speed. Assuming a model that uses Fresnel diffraction theory, two Fourier transforms are needed to simulate the propagation of light between individual measurement planes. The results of our numerical study for the number and location of sensing planes, as well as spatial sampling, may therefore be used to place requirements on real-time computing hardware in order to keep pace with atmospheric turbulence. A possible further optimization could be the use of other algorithms for reconstruction beyond the nominal Gerchberg-Saxton technique.\cite{Oliker_18,Fienup:06,Brady:06} These may offer faster convergence, although it is expected optimal measurement plane performance criteria would remain broadly unchanged as this is set by diffraction and turbulence strength effects.

\subsection*{Disclosures}

JRC has filed a non-provisional patent related to optical designs for the sensor studied in this article. The other authors have no relevant financial interests in the manuscript and no other potential conflicts of interest to disclose.

\subsection*{Code, Data, and Materials Availability} 
Data from plots are available from the authors upon request. Numerical simulations in this article used publicly available Matlab scripts developed by J.D. Schmidt, as well as scripts from the commercially-available WaveProp and AOTools Matlab packages.

\subsection*{Acknowledgements}

This research was supported in part by the Air Force Office of Scientific Research (AFOSR) grant number FA9550-22-1-0435. We acknowledge support from Northrop Grumman Space Systems. A preliminary and limited subset of this work was previously included in the Proceedings of SPIE Astronomical Telescopes + Instrumentation, 2022.\cite{letchev_22} We also thank Mark Spencer for helpful discussions regarding spatial sampling and wavefront sensing in general.

%%%%% References %%%%%

\bibliography{Assessing_Phase_Reconstruction_Accuracy_for_Different_Nonlinear_Curvature_Wavefront_Sensor_Configurations}
\bibliographystyle{spiejour}

\end{spacing}
\end{document}